\documentclass[aps,prl,twocolumn,tightenlines,superscriptaddress]{revtex4}

\usepackage{amssymb}
\usepackage{graphicx}
\usepackage{todonotes}
\usepackage{amsmath}
\usepackage{amsbsy}
\usepackage{amsthm}
\usepackage{bbm}
\usepackage{epsfig}

\newcommand{\tr}{\text{Tr}}

\newcommand{\be}{\begin{equation}}
\newcommand{\ee}{\end{equation}}
\newcommand{\beq}{\begin{eqnarray}}
\newcommand{\eeq}{\end{eqnarray}}

\usepackage{xcolor}

\begin{document}

\title{Contextuality and Nonlocality in Decaying Multipartite Systems}
\author{Beatrix C. Hiesmayr}
\email{Beatrix.Hiesmayr@univie.ac.at}
\affiliation{Faculty of Physics, University of Vienna, Boltzmanngasse 5, 1090 Vienna, Austria}
\author{Jan-\AA ke Larsson}
\email{jan-ake.larsson@liu.se}
\affiliation{Department of Electrical Engineering, Link{\"o}ping University,
  58183 Link\"oping, Sweden}

\begin{abstract}
  Everyday experience supports the existence of physical properties
  independent of observation in strong contrast to the predictions of quantum
  theory. In particular, existence of physical properties that are independent
  of the measurement context 
  is prohibited for certain quantum systems. This property is known as \emph{contextuality}. 
  This paper studies whether the process of decay in space-time generally destroys the ability of
  revealing contextuality. We find that in the most general situation
  the decay property does not diminish this ability. However, applying certain
  constraints due to the space-time structure either on the time evolution of
  the decaying system or on the measurement procedure, the criteria revealing
  contextuality become inherently dependent on the decay property or an impossibility. In
  particular, we derive how the context-revealing setup known as Bell's
  nonlocality tests changes for decaying quantum systems. Our findings
  illustrate the interdependence between hidden and local hidden parameter
  theories and the role of time.
\end{abstract}

\pacs{03.65.Ud, 03.65.Yz, 03.65.Ta}

\maketitle


\textit{Introduction.---} The notion of (non)-contextuality has its origins in
logic of non-simultaneously decidable propositions~\cite{Specker} and has been
extensively studied, in particular with respect to the question of the
existence of hidden parameters~\cite{KochenSpecker} and in terms of
applications such as being the key property for a computation speed up in
quantum algorithms~\cite{Larsson,Raussendorf,Howard}. A
theory is said to be
non-contextual if every random variable only depends on the choice of
the measurement but not on the choice of other compatible measurements that are
co-measured,  its \emph{measurement context}. If this independence condition
does not hold, we call it contextual. This property can be tested through
criteria designed such that they distinguish these two cases given the
conditions, e.g.~\cite{KochenSpecker,Peres,Mermin1,Klyachko}. Another way to
formulate this is to view measurements in groups that are compatible to each
other, contexts, as having outcomes that are jointly distributed within each
context but stochastically unrelated between contexts. In quantum mechanics
different contexts correspond to different mutually incompatible conditions,
so no stochastic relation is present. The question is whether a joint
distribution on the full set of observables exists, then providing a
non-contextual model, or if it does not exist, so that the system can be said
to be \emph{contextual}.

This paper considers decaying quantum systems and asks whether the process of
decay diminishes or destroys the contextual feature present at a certain time
point. In particular, we will consider the question whether for a set of
measurements, the impossibility of pre-determined outcomes holds for all times
if it holds for a certain time in the past (or future). This is a non-trivial
question since decaying systems live in Hilbert spaces that have to be
separated in a ``\textit{surviving part}'' and a ``\textit{decaying part}'',
where only the surviving part is available for the intended measurements. In
particular, the all-important choice of context is only possible for the
surviving part.

The paper is organized as follows. First, we stress that there are two
different types of dichotomic measurements for decaying (multipartite)
systems. We then show that in the joint-particle measurement scenario (defined
below) every criterion revealing contextuality can be turned into a criterion
that is violated (revealing contextuality) for all times if it is violated at
a time point in the past (or future). This proves that the property of
contextuality, the impossibility to pre-assign results to a measurement,
persists in time, i.e., is unaffected by the decay property. In the case of
single-particle measurements (defined below), which are the most common
experimental situations, we show that the conditions of compatibility are more
involved. Last but not least we elaborate how the specific contextuality test known as
``Bell's nonlocality'' leads to Bell inequalities for decaying systems. This
in particular illuminates how dynamical nonlocality differs with respect to
stable systems.

\begin{figure}
  \includegraphics[width=0.49\textwidth,keepaspectratio=true]{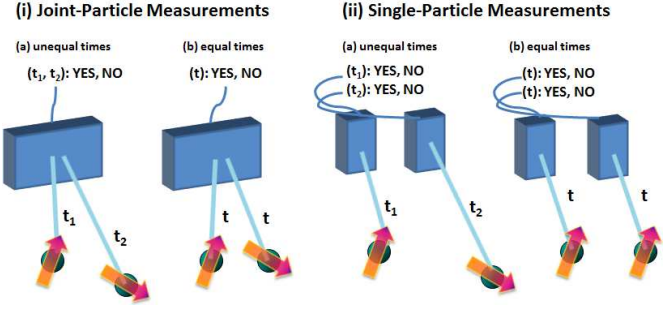}
  \caption{Schematic view of the two dichotomic questions for a two particle
    scenario. Note that the two particles can be in a separable or entangled
    state.}\label{jointsinglemeasurements}
\end{figure}

\textit{Two distinct dichotomic measurements on a multipartite decaying
  system.---} A decaying system has a natural separation into a
``\textit{surviving part}'' and into a ``\textit{decaying part}'' whose
Hilbert spaces are disjoint. The crucial point is that any experimental setup
only has access to the surviving part. Consequently, there exists two
dichotomic inequivalent information complete questions that can be raised to an $n$-partite
decaying system:
\begin{enumerate}
\item[(i)]{\it Joint-particle measurements}: Is the decaying system in the
  state $|\psi\rangle=\sum_{i=1}^{d_1\cdot d_2\cdot\dots\cdot d_n}
  \alpha_{i}|e^{(i)}\rangle$ at time $t_1,t_2,\dots,t_n$ or not?
\item[(ii)]{\it Single-particle measurements}: Is the decaying system in the
  state $|\phi_1\rangle=\sum_{i=1}^{d_1} \alpha_{i}|f_1^{(i)}\rangle$ for
  particle $1$ at time $t_1$ or not, in the state
  $|\phi_2\rangle=\sum_{i=1}^{d_2} \beta_{i}|f_2^{(i)}\rangle$ at time $t_2$
  for particle $2$ or not, \dots, and in the state
  $|\phi_n\rangle=\sum_{n=1}^{d_2} \gamma_{i}|f_n^{(i)}\rangle$ at time $t_n$
  for particle $n$ or not?
\end{enumerate}
Here we have assumed that the decaying systems consist of $n$ particles
($n=1,2,\dots$) each described by $d_n$ degrees of freedom. The vectors
$|e^{i}\rangle, |f^{i}_j\rangle$ form an orthogonal basis of the surviving
part of the Hilbert space, respectively. These two conceptually different
measurement procedures and their two different cases (equal and unequal times)
are illustrated in Fig.~\ref{jointsinglemeasurements}.

\textit{Time evolution.---} Since the decay is a Markov process we can model
the system as an open quantum system (for applications see e.g. Refs.~\cite{Bernabeu,Smolinski}). As shown in
Ref.~\cite{HiesmayrOpenQuantumSystem} any decaying multipartite system can be
modeled by a Hamiltonian $H$ covering the surviving part $s$ and a Lindblad
operator $L$ connecting the $s$ part with the decaying part $d$, i.e.  \beq
\mathcal{H}=|s\rangle\langle s|\otimes H\quad\textrm{and}\quad
\mathcal{L}=|d\rangle\langle s|\otimes L\;, \eeq which satisfies the Lindblad
master equation~\cite{Lindblad1,Lindblad2} ($\hbar\equiv 1$)
\begin{equation}\label{Lindbladequation}
  \frac{\mathrm{d}\rho}{\mathrm{d}t} =
  -\,i [\mathcal{H}, \rho]-\frac{1}{2}\sum \left(\mathcal{L}_i^\dagger \mathcal{L}_i\,\rho+\rho\, \mathcal{L}_i^\dagger \mathcal{L}_i-2 \mathcal{L}_i\,\rho\, \mathcal{L}_i^\dagger\right) \,,
\end{equation}
where $\rho$ is the state of the decaying system and is divided into
a ``surviving'' and a ``decaying'' part: $\rho=|s\rangle\langle
s|\otimes\rho_s+|d\rangle\langle s|\otimes\rho_d$. Obviously, the decaying
part has to be determined by the time evolution of the surviving part, i.e.
\beq \rho_{d}(t)=L\,\int_0^t\rho_{s}(t')dt'\,L^\dagger\;, \eeq and the decay
rate $\Gamma$ is given by $L^\dagger L=\Gamma$. The solution
$\rho(t)=|s\rangle\langle s|\otimes\rho_s(t)+|d\rangle\langle
s|\otimes\rho_d(t)$ of this differential equations can be derived for any
number of particles and is referred to as a ``\textit{joint-particle}'' time
evolution.

To obtain a ``\textit{single-particle}'' time evolution of $n$ particles we
have to exploit the usual tensor product structure for the Hamiltonian and the
generators of the decay \beq
\lefteqn{H_\mathrm{single-particle}=}\nonumber\\
&&H\otimes\mathbbm{1}^{\otimes n-1}+\mathbbm{1}\otimes H\otimes\mathbbm{1}^{\otimes n-2}+\dots+\mathbbm{1}^{\otimes n-1}\otimes H\nonumber\\
\lefteqn{L_\mathrm{single-particle}=}\nonumber\\
&&L\otimes\mathbbm{1}^{\otimes n-1}+\mathbbm{1}\otimes
L\otimes\mathbbm{1}^{\otimes n-2}+\dots+\mathbbm{1}^{\otimes n-1}\otimes L\;.
\eeq Note that in this case the total state is divided for two particles into
four subspaces, surviving-surviving (ss), surviving-decaying (sd),
decaying-surviving (ds) and decaying-decaying (dd) and defined for two
different times. Explicit solutions for both cases are discussed later.

\textit{Revealing Contextuality in Decaying Systems and as given by Space-Time
  structure.---} We start by exploiting the state-dependent
Klyachko-Can-Binicio\u{g}lu-Shumovsky-inequality \cite{Klyachko} that works
for any system of dimension three or larger. It is given by \beq\label{KCBS}
I_{KCBS}&=&\tr(O_1 O_2 \rho)+\tr(O_2 O_3 \rho)+\tr(O_3 O_4 \rho)\nonumber\\
&&+\tr(O_4 O_5 \rho)+\tr(O_5 O_1 \rho)\;\geq\; -3 \;,\eeq where each pair of
observables has to be compatible (which means for quantum mechanics that the
observables are orthogonal, i.e. $\tr O_i O_{\text{mod}(i+1,5)}=0$). For an
optimal choice of quantum observables with respect to some given pure state
$\rho$ it is known that the quantum bound $5-4\sqrt5\approx-3.944$ can be
reached which, consequently, reveals the contextual feature of quantum
mechanics. Since the operators $O_i$ do not need to have a tensor-product
structure they generally correspond to ``joint-particle'' measurements, type
$(i)$, and the relevant time evolution is a joint-particle time evolution. Let
us assign the numbers $+1$ to a YES outcome and $-1$ to a NO outcome,
obviously the physics does not depend on that choice (we will exploit this
fact later). Any expectation value can be rewritten to only depend on the
surviving part through \beq
\tr O_i \rho &=&\tr (2 \mathcal{P}_i-\mathbbm{1})\rho=\tr (2 P_i-\mathbbm{1})\rho_s-\tr\rho_d\nonumber\\
&=&\tr (2 P_i-\mathbbm{1})\rho_s-(1-\tr\rho_s)\;, \eeq where $\mathcal{P}_i$
is a projector on the full space and $P_i$ the corresponding projector onto
the surviving part (note that no projection onto the decaying part is
possible). If we assign instead the numbers $-1$ to a YES outcome and $+1$ to
a NO outcome, we obtain an overall minus sign, but if we assign this
relabelling to the projector $P_i$ only onto the decaying part, we obtain a
relative sign change. This situation corresponds to two physical distinct
questions that are identical for stable systems, i.e., (here we assume for
simplicity that all particles are measured jointly at the same time instance)
\begin{itemize}
\item[(A)] \textit{Is the system in the state $|\psi_i\rangle$ at time $t$ or
    not?}: $\tr O_i \rho(t)=\tr (2 P_i-\mathbbm{1})\rho_s(t)-(1-\tr\rho_s(t))$
\item[(B)] \textit{Is the system not in the state $|\psi_i^\perp\rangle$ with
    $\langle \psi|\psi^\perp\rangle=0$ at time $t$ or is it?}: \\ $\tr \bar{O}_i
  \rho(t)=\tr (2 P_i-\mathbbm{1})\rho_s(t)+(1-\tr\rho_s(t))$
\end{itemize}
The first question outputs $+1$ if the system is in the state
$|\psi_i\rangle$, while the second question outputs $+1$ if the system is in
the state $|\psi_i\rangle$ or if it has decayed.  In a measurement of $O_iO_j$
there are now two possibilities depending on whether we choose the same or
different  assignments of $+1$ and $-1$ to
the measurement outcomes, i.e.:
\beq
\tr O_1 O_2 \rho=\tr (2 P_1-\mathbbm{1})(2 P_2-\mathbbm{1})\rho_s\pm(1-\tr\rho_s)
\eeq

Inserting these expectation values into $I_{KCBS}$ we obtain an inequality for
decaying subsystems that reads
\beq
\begin{split}
  \label{KCBSdecay}
  I_{KCBS}^{\rm{decay}}(t)&=\tr((2 P_1-\mathbbm{1}) (2 P_2-\mathbbm{1})
  \rho_s(t))\\
  &\quad+\tr((2 P_2-\mathbbm{1}) (2 P_3-\mathbbm{1}) \rho_s(t))\\
  &\quad+\tr((2 P_3-\mathbbm{1}) (2 P_4-\mathbbm{1}) \rho_s(t))\\
  &\quad+\tr((2 P_4-\mathbbm{1}) (2 P_5-\mathbbm{1}) \rho_s(t))\\
  &\quad+\tr((2 P_5-\mathbbm{1}) (2 P_1-\mathbbm{1}) \rho_s(t))\\
  &\quad+ c\cdot(1-\tr\rho_s(t))\;\geq\;-3\,.\label{eq:1}
\end{split}
\eeq

Due to the freedom of assigning $+1$ and $-1$ to the measurement outcomes one
can control the additional term $c\cdot(1-\tr\rho_s)$. The optimum is reached
by choosing alternating assignments of $-1$ and $+1$ to the event of finding
that the system has decayed, resulting in $c_{\textrm{opt}}=-3$. Since
$\rho_s$ vanishes with increasing time $t$ the inequality
$I_{KCBS}^{\textrm{decay}}(t\longrightarrow\infty)$ approaches the classical
bound $-3$ from below. Consequently, we have shown that if a decaying system
violates this criterion at a given point in time, the violation decreases as
time goes on but will remain for all times, thus the contextual feature
remains. Note that this result holds only for joint-particle measurements and
corresponding joint-particle time evolutions as we will discuss later in
detail.

Let us consider another inequality revealing contextuality, the well known
Mermin-Peres square~\cite{Mermin1,Peres}, which is known to be
state-independent \beq
I_{MP}&=&\tr \left\lbrace (A_{11}A_{12}A_{13}+A_{21}A_{22}A_{23}+A_{31}A_{32}A_{33}\right.\nonumber\\
&&\left.+A_{11}A_{21}A_{31}+A_{21}A_{22}A_{23}-A_{31}A_{32}A_{33})
  \rho\right\rbrace\leq 6\nonumber \eeq with \beq
(A)_{ij}&=&\left(\begin{array}{ccc}\sigma_x\otimes\mathbbm{1}&\mathbbm{1}\otimes\sigma_z&\sigma_x\otimes\sigma_z\\
    \mathbbm{1}\otimes\sigma_x&\sigma_z\otimes\mathbbm{1}&\sigma_z\otimes\sigma_x\\
    \sigma_x\otimes\sigma_x&\sigma_z\otimes\sigma_z&\sigma_y\otimes\sigma_y\end{array}\right)_{ij}\;.
\eeq It involves the product of three operators (being measured jointly!) and that all products compute.
For decaying quantum systems we obtain \beq
\tr O_1 O_2 O_3 \rho &=&\pm \tr (2 P_1-\mathbbm{1})(2 P_2-\mathbbm{1})(2 P_3-\mathbbm{1})\rho_s(t)\nonumber\\
&&-(1-\tr\rho_s(t))\;, \eeq where we obtained again a relative sign depending
on our assignment of $+1$ or $-1$ to a ``YES'' event. Thus the Mermin-Peres
version for decaying systems (for both sign choices) becomes \beq
I_{MP}^\textrm{decay}=6 \tr \rho_s(t)+4 (1-\tr \rho_s(t))= 2
\tr\rho_s(t)+4\leq 4\;,\nonumber \eeq which is obviously violated for any
initial state and for all times.

Straightforwardly, one can also optimize the corresponding contextuality criteria for more than two particles, e.g. the state independent criterion for three qubit systems introduced in Ref.~\cite{Mermin2} becomes
\beq
I_{\textrm{3 particles}}^{\textrm{decay}}
&=&3+2 \tr\rho_{sss}+2 \tr\rho_{sds}+2 \tr\rho_{dsd}\leq 3\;.\nonumber
\eeq
Again in the limit of infinite time, we approach the bound from above showing that if contextuality can be witnessed by this inequality for a certain time instance, then it holds for all times.

\begin{figure}
\includegraphics[width=0.41\textwidth, height=0.25\textwidth]{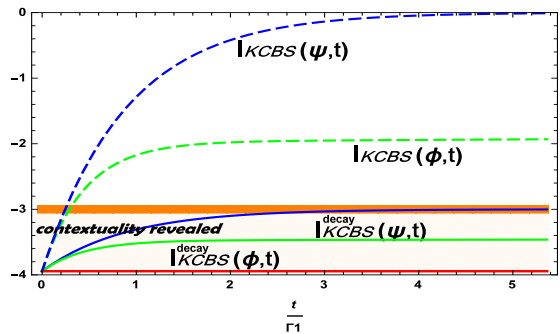}
\caption{The curves show the KCBS-inequality optimized over all five observables at each time point for entangled neutral K-meson pairs (in units of time life of the shortest decay rate). The blue curves correspond to initial Bell states $|\psi^\pm\rangle\equiv \psi$ and the green curves for $|\phi^\pm\rangle\equiv\phi$ given in the basis of the eigenstates of the Hamiltonian (the mass-eigenstates). The dashed curves are the results of the unoptimized version, Ineq.~(\ref{KCBS}), and the bold curves of the optimized version, Ineq.~(\ref{KCBSdecay}). For longer time scales also $I_{\textrm{KCBS}}^{\textrm{decay}}(\phi,t)$ approaches from above the classical bound $-3$.}\label{figureKCBS}
\end{figure}

\textit{Refining The Contexts By The Space-Time-Structure.---}
The simplest decaying quantum system is a two-state system (qubit). The solution of the Lindblad equation~(\ref{Lindbladequation}) in terms of Kraus operators  $Q_i$ and assuming two decays constants $\Gamma_{1,2}$ and two energies $E_{1,2}$ is given by
\beq
\rho(t)=\sum_{i=s}^{d}Q_i(t)\rho_s Q_i^\dagger(t)
\eeq
with
$
Q_s(t)=|s\rangle\langle s|\otimes K_s(t)$
, $Q_d(t)=|d\rangle\langle s|\otimes K_d(t)
$
, $K_s(t)=\textrm{diag}\{e^{-\frac{\Gamma_1+i (E_2-E_1)}{2} t},e^{-\frac{\Gamma_2-i(E_2-E_1)}{2} t}\}$ and $K_d(t)=\textrm{diag}\{\sqrt{1-e^{-\Gamma_1 t}},\sqrt{1-e^{-\Gamma_2 t}}\}$. Obviously both discussed criteria for contextuality can not be violated since at least a three dimensional system for the KCBS-criterion or a four dimensional system for the Mermin-Peres-criterion is required. Therefore we proceed to bipartite identical two-state systems. The joint-particle time evolution in terms of Kraus operators is derived to
\beq
\rho(t)=\sum_{i=1}^{2}Q_i(t)\rho_{ss} Q_i^\dagger(t)
\eeq
with $Q_1(t)=|s\rangle\langle s|\otimes K_s(t)\otimes K_s(t)$, $Q_2(t)=|d\rangle\langle s|\otimes K_{dd}(t)$
and \beq
K_{dd}(t)&=&\textrm{diag}\{\sqrt{1-e^{-2\Gamma_1 t}},\sqrt{1-e^{-(\Gamma_1+\Gamma_2) t}},\nonumber\\
&&\hphantom{\textrm{diag}\{} \sqrt{1-e^{-(\Gamma_1+\Gamma_2) t}},\sqrt{1-e^{-2\Gamma_2 t}}\}\;.\nonumber\eeq
Note that in the decay-decay (dd) part the tensor-product structure in the time parameter is lost.

We can now use  the above two criteria for contextuality, KCBS and Mermin-Peres. Fig.~\ref{figureKCBS} shows the result for the flavor-oscillating and decaying K-mesons system for the $KCBS$ criterion ($E_1-E_2\equiv \Delta m=m_2-m_1=3.5\cdot 10^{-12}MeV$ and $\Gamma_1\approx 2 \Delta m\approx 600\Gamma_2$). Let us remark that the behaviour of the violation depends strongly on the initial Bell state (symmetric or antisysmmetric) showing an additional state dependence due to the decay property. Initial entangled states for pairs of K-mesons can be produced~\cite{CPLEAR,KLOE}, however, it is not clear how joint-particle measurements may be technically realized. A suitable system for the application of the Mermin-Peres criterion are spin entangled hyperon-antihyperon systems which also decay via weak interactions but have half-integer spins as discussed in Ref.~\cite{HyperonHiesmayr}.

Typically in decaying bipartite systems one assumes independent time evolutions for the individual particles. The solution of the Lindblad equation~(\ref{Lindbladequation}) has then to be separated into the four parts $(ss),(sd), (ds)$ and $dd$, i.e. we obtain a state conditioned to the two time choices $t_l,t_r$ ($l$\dots left, $r$\dots right)
\beq
\rho(t_l,t_r)\equiv\sum_{i,j=s}^d Q_{ij}(t_l,t_r)\;\rho_{ss}\; Q_{ij}^\dagger(t_l,t_r)
\eeq
with
$Q_{ij}(t_l,t_r)=|ij\rangle\langle ss|\otimes K_i(t_l)\otimes K_j(t_r)
$.
Consequently, the expectation value of two jointly measured observables becomes
\beq
\tr\{O_1 O_2\rho\}&=&\tr\{(2 \mathcal{P}_1-\mathbbm{1})(2 \mathcal{P}_2-\mathbbm{1})\rho(t_l,t_r)\}\nonumber\\
&=&\sum_{j,k=s}^{d}\tr\biggl\lbrace(2 (P_1)^{jk}-\mathbbm{1})(2 (P_2)^{jk}-\mathbbm{1})\nonumber\\
&&\quad\qquad Q_{jk}(t_l,t_r)\;\rho_{ss}\;Q_{jk}^\dagger(t_l,t_r)\biggr\rbrace
\eeq
where we have for joint-particle measurements that $P^{jk}$ is a projector $P$ on the $ss$ part and else the unity operator.
Since we can construct again commuting operators we can apply also in this case the contextuality criteria revealing the contextual nature even for two different times (case (i)(a) of Fig.~\ref{jointsinglemeasurements}). On the other hand, if we perform single-particle measurements, the compatibility of the operators cannot be obtained, consequently we cannot apply the criteria. One may think that increasing the
number of particles (like in Ref.~\cite{Mermin2}) or increasing the number of observables (like in Ref.~\cite{Cabello}) may help, however, it is principally not possible to restore the compatibility. A context can only be generated for single particle measurements if Bell's locality assumption in space-time is taken in consideration.

\textit{Connection to Bell's theorem.---} As is well known if one reduces the
number of measurements for the KCBS or Mermin-Peres contextuality test and
assumes a tensor product structure of the involved observables, one obtains
the $Bell-CHSH$ inequality~\cite{chsh}. The crucial point here is that by Bell's locality assumption one implies indirectly individual particles propagating in space-time. Still there are the two options of measurements. For
joint-particle measurements on a bipartite system we obtain
\beq
\label{Bellrenomalized} -2 \tr\rho_{ss}(t)\leq \tr
\widehat{\textrm{Bell}}\rho_{ss}(t)\leq 2 \tr\rho_{ss}(t)\;,
\eeq
which is violated for all times for any initial state that violates the Bell
inequality since $\rho_{ss}(t)/\tr\rho_{ss}(t)$ is a normalized
state. This inequality is a contextuality proof, but no test
for Bell's locality hypothesis since both particles are measurement jointly.

Bell's locality hypothesis requires individual particles located at different locations in space-time imposing single-particle measurements and single-particle time evolution. In this case the single-particle measurements do depend on the time choices of $t_l,t_r$ and the projections, i.e. the operators under investigation become time dependent
\beq
\tilde{O}&=&\sum_{j,k=s}^{d} 2 K_j^\dagger(t_l)(P^l)^j K_j(t_l)\otimes K_k^\dagger(t_r)(P^r)^k K_k(t_r)\nonumber\\
&&-K_j^\dagger(t_l) K_j(t_l)\otimes K_k^\dagger(t_r) K_k(t_r)
\eeq
with
\beq
(P)^j&=&\left\lbrace\begin{array}{c} j=s:P\\
j=d: \mathbbm{1}\end{array}\right.\;.
\eeq
These operators are always commuting (compatible) since they have the natural
context of being measured at different instances in space-time. Note in particular,
in the case all operators $P^l,P^r$ are chosen to be the same, we obtain a
nontrivial Bell inequality violated by different time choices (distances from
the source), exhibiting a kind of ``\textit{dynamical}'' nonlocality. Such a type
of Bell's inequality being experimentally feasible with a further trick was
introduced for entangled decaying K-mesons in Ref.~\cite{Hiesmayr:2012}.

\textit{Conclusions.---} The contextual property is conjectured to be key
ingredient of quantum theory. We discuss how this property can be
revealed in decaying quantum systems under the assumption that the entire time evolution including the decay property is independent of the measurement
choices.  We found that any criterion based on
joint-particle measurements and joint-particle or single-particle time evolution can be rewritten
to display the contextual nature, in principle at any instant in time. That proves that the decay property per se is not sensitive to the notion of measurement contextuality. 

Interestingly, we find that the standard contextuality criteria can not be applied when we assume single-particle measurements, because the compatibility requirement is not fulfilled. The requirement can be restored by generating the context via assumption of space-time-localization leading to state-dependent \textbf{and} decay property dependent Bell-like inequalities.

This findings prove the crucial difference between assigning hidden parameters to measurement outcomes and local hidden parameters to the involved
state; such as that the context is achieved by different requirements on the setup:
compatible joint-particle measurements or localization assumption in space-time. It illustrates the foundational different concepts of \textit{time} in time evolutions of states and in space-time with respect to compatible measurement setups.

\textbf{Acknowledgement:} B.C. Hiesmayr acknowledges gratefully the Austrian
Science Fund (FWF-P26783). Both authors want to thank the COST-action
MP1006 ``\textit{Fundamental Problems in Quantum Physics}'' that initiated
this work.

\end{document}